\documentclass[manuscript]{acmart}
\usepackage{multirow}
\usepackage{soul} 
\usepackage{enumitem}
\usepackage{array}

\newcommand{\red}[1]{\textcolor{black}{#1}}
\newcommand{\black}[1]{\textcolor{black}{#1}}
\usepackage{soul}
\setstcolor{red}
\AtBeginDocument{%
  \providecommand\BibTeX{{%
    \normalfont B\kern-0.5em{\scshape i\kern-0.25em b}\kern-0.8em\TeX}}}

\setcopyright{rightsretained}
\acmJournal{PACMHCI}
\acmYear{2022} \acmVolume{6} \acmNumber{CSCW1} \acmArticle{75} \acmMonth{4} \acmPrice{}\acmDOI{10.1145/3512922}

\begin{document}

\title{An Exploration of Captioning Practices and Challenges of Individual Content Creators on YouTube for People with Hearing Impairments}


\author{Franklin Mingzhe Li}
\affiliation{
  \institution{Carnegie Mellon University}
  \city{Pittsburgh}
  \state{Pennsylvania}
  \country{USA}
}
\email{mingzhe2@cs.cmu.edu}

\author{Cheng Lu}
\affiliation{
  \institution{University of Toronto}
  \city{Toronto}
  \state{Ontario}
  \country{Canada}
}
\email{ericlu.lu@mail.utoronto.ca}

\author{Zhicong Lu}
\affiliation{ 
  \institution{City University of Hong Kong}
  \city{Hong Kong}
  \country{China}
}
\email{zhiconlu@cityu.edu.hk}

\author{Patrick Carrington}
\affiliation{%
    \institution{Carnegie Mellon University}
    \city{Pittsburgh}
    \state{Pennsylvania}
    \country{United States}
 }
 \email{pcarrington@cmu.edu}

\author{Khai N. Truong}
\affiliation{
  \institution{University of Toronto}
  \city{Toronto}
  \state{Ontario}
  \country{Canada}
}
\email{khai@cs.toronto.edu}

\renewcommand{\shortauthors}{Li et al.}
\begin{abstract}
Deaf and Hard-of-Hearing (DHH) audiences have long complained about caption qualities for many online videos created by individual content creators on video-sharing platforms (e.g., YouTube). However, there lack explorations of practices, challenges, and perceptions of online video captions from the perspectives of both individual content creators and DHH audiences. In this work, we first explore DHH audiences' feedback on and reactions to YouTube video captions through interviews with 13 DHH individuals, and uncover DHH audiences' experiences, challenges, and perceptions on watching videos created by individual content creators (e.g., manually added caption tags could create additional confidence and trust in caption qualities for DHH audiences). We then discover individual content creators' practices, challenges, and perceptions on captioning their videos (e.g., back-captioning problems) by conducting a YouTube video analysis with 189 captioning-related YouTube videos, followed by a survey with 62 individual content creators. Overall, our findings provide an in-depth understanding of captions generated by individual content creators and bridge the knowledge gap mutually between content creators and DHH audiences on captions.
\end{abstract}



\begin{teaserfigure}
  \includegraphics[width=\textwidth]{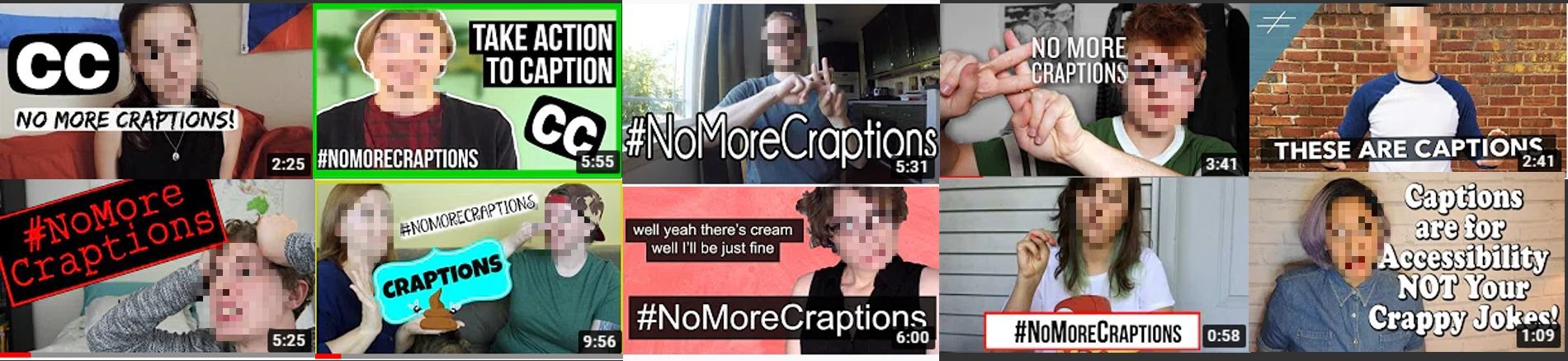}
  \caption{YouTubers' and Deaf and Hard-of-Hearing audiences' complaints and campaigns on asking for better captions of videos created by individual content creators on video-sharing platforms (e.g., \#NoMoreCraptions).}
  \Description{Ten thumbnails of YouTubers' and Deaf and Hard-of-Hearing audiences' complaints and campaigns on asking for better captions of videos created by individual content creators on video-sharing platforms (e.g., \#NoMoreCraptions).}
  \label{fig:teaser}
\end{teaserfigure}
\maketitle

\section{Introduction}

\black{People who are Deaf and Hard-of-Hearing (DHH) consume a large volume of videos from video-sharing platforms (e.g., YouTube) for various reasons, such as entertainment, relaxation, and learning new skills. There is an imperative need for DHH individuals to have accessible video captions to understand video content. As of 2012, over 37 million people in the U.S. have hearing impairments \cite{blackwell2014summary} and often require captioning services to access audio and audio-visual information.} 


\black{To caption videos, both human captioning approaches and automatic captioning approaches exist. Video captioning through human effort is time-consuming and inexperienced users usually spend at least six to eight times the length of the video to create captions from scratch \cite{ClosedCa43:online}.} Individual content creators on YouTube upload their videos, usually once per day or multiple times a week. Most of these content creators do not entirely rely on the income from their uploaded videos for a living, and 96.5\% of YouTube content creators do not even make it above the U.S. poverty line with the income from being a content creator \cite{Success63:online}. Many of the popular YouTubers must still maintain their primary employment elsewhere \cite{WhySomeS82:online}. Such individual content creators tend to have limited time or budget for their videos. Therefore, this special situation and background of individual content creators may affect the caption quality and DHH audiences' experiences with watching videos uploaded from individual content creators. \black{To save time for captioning, some content creators use Automatic Speech Recognition (ASR) technology to caption videos \cite{Useautom75:online}. However, the current ASR technology is error-prone due to variability and complexity of speech (e.g., background noise, accent), which may make the video content confusing to audiences \cite{hazen2006automatic,benzeghiba2007automatic,glasser2017deaf,o2008automatic,araujo2014context}}.

In the past few years, many content creators and DHH \black{viewers have} started multiple campaigns and events to call for better captions of online videos (Fig. \ref{fig:teaser}), for example, the ``\#NoMoreCraptions'' campaign \cite{1NewMess92:online,ANewCamp15:online} initiated by Rikki Poynter \cite{RIKKIPOY3:online}, a deaf content creator on YouTube. Poynter commented on her inspiration of starting the campaign and complained about the quality of YouTube captions, ``\textit{The lack of proper closed captioning has always inspired me, but I think the final straw was when I saw viewers of big YouTubers add in unnecessary commentary, jokes, etc. to closed captioning on videos and then yelling at the d/Deaf/HOH community when they got called out for it. I just wanted to try to make something big happen. I wanted help. I needed help.}'' \cite{Autogene63:online}. After the ``\#NoMoreCraptions'' campaign initiated, many content creators and DHH viewers supported the campaign by filming videos to express the importance of captions (e.g., Fig. \ref{fig:teaser}) and posting their perspectives on social platforms (e.g., Twitter \cite{nomorecr8:online}). Although prior research explored DHH audiences feedback towards captions in general (e.g., \cite{kawas2016improving}), little research has been done to explore the barriers of DHH audiences' experiences and challenges with video captions created by individual content creators. Furthermore, prior research mostly focused on consumers' perspectives (e.g., DHH audiences) of captions, which lacks explorations on the perspectives from caption producers (e.g., individual content creators). To improve the DHH audiences' experiences with captions created by individual content creators, we need to mutually understand DHH audiences' feedback on video captions produced by individual content creators on video-sharing platforms and individual content creators' practices, perceptions and challenges in captioning. In our work, we are interested in exploring the following research questions:

\begin{itemize}
    \item (RQ1) From the perspective of DHH audiences, what are the experiences with, feedback towards, and potential improvements of the captions provided by individual content creators?
    
    \item (RQ2) From the perspective of individual content creators, what are the existing practices and perceptions of captioning their own videos?
    
    \item (RQ3) From the perspective of individual content creators, what are the challenges and problems involved in captioning their own videos?
\end{itemize}

\black{To explore RQ1, we first conducted semi-structured interviews with 13 DHH YouTube viewers. Based on the feedback, we revealed the practices, perceptions, and challenges of DHH audiences in video searching and filtering, concerns of caption quality, and potential improvements (e.g., small talking head at the corners of the screen made DHH audiences hard to read the lips (Fig. \ref{fig:lipreading})), and their reactions and mitigation strategies to poor-caption problems. To understand RQ2 and RQ3, we then conducted a video analysis with 189 non-commercial YouTube videos uploaded by individual content creators and a survey study with 62 individual content creators. We presented the existing captioning methods for individual content creators to caption their videos and their perceived benefits and drawbacks of these methods (e.g., caption errors could potentially demonetize the video). We further explored the content creators' perceptions and challenges of captioning (e.g., back-captioning all previous videos that do not have captions due to knowledge gaps and lack of awareness). Finally, we extended our findings through the discussion of developing a DHH-friendly video recommendation system, minimizing the effort on previewing caption quality, improving and motivating high-quality community captions, presenting different levels of caption details, and opportunities with lip-reading. 
Overall, we believe our findings provide an in-depth understanding of captions generated by individual content creators and bridge the gap between DHH audiences and individual content creators on creating, interpreting, and presenting captions of user-generated videos online. Our findings will also shed light on future online-captioning system design (e.g., community-contributed captions) to HCI and CSCW researchers by understanding the practices and challenges of captions from different perspectives among caption creators and consumers.}

\section{Background and Related Work}
\subsection{\black{Usefulness of Online Video Captions for DHH Individuals}}
The large volume of videos on social media platforms and online learning environments, such as Massive Open Online Courses (MOOCs), now enable people to obtain comprehensive information online (e.g., \cite{li2021non}). Shiver and Wolfe \cite{shiver2015evaluating} interviewed 20 DHH individuals and found that all of them expressed the importance and usefulness of having captions when watching online videos with captions. Shiver and Wolfe \cite{shiver2015evaluating} further mentioned that even machine-generated captions that contain errors could also help DHH individuals understand video content. Beyond enabling DHH audiences to have access to video contents, captions could also benefit other groups of people, such as second-language learners \cite{collins2013using,garza1991evaluating,markham2001influence,neuman1992captioned,vanderplank1988value,whitney2019captioning} and native hearing audiences \cite{gernsbacher2015video}. Furthermore, having videos captioned could potentially benefit SEO (Search Engine Optimization) and help the video soar in searching \cite{WhatIsVi84:online}.

Although Federal Communications Commission (FCC) has enforced online videos to be captioned through legislation \cite{National23:online}, many DHH \black{individuals} have reported and started multiple events and campaigns to call for better captions for online videos created by individual content creators \cite{1NewMess92:online,ANewCamp15:online}. \black{To understand DHH audiences' perceptions on caption qualities, Kawas et al. \cite{kawas2016improving} conducted co-design workshops with DHH individuals and found that their participants struggled with the low-quality captions. Beyond low-quality captions, Tyler et al. \cite{tyler2009effect} found that the rate of caption delivery affects the comprehension of contents. Furthermore, the variability and complexity of human speech may affect the accuracy of captions \cite{kushalnagar2014accessibility}. More specifically, having background noise, different speech rate or speaking styles would affect the quality of the caption, thus affect the comprehension and understanding of captions \cite{o2008automatic}.}


\subsection{\black{Existing Captioning Methods}}

There exist several approaches to captioning videos, including manually captioning, automatic captioning, and crowd-sourced captioning. To caption videos manually, video producers could generate captions through captioning softwares by syncing a script with time points in a video. This approach usually takes four to six times the video's length to create the captions by professional captionists \cite{ClosedCa43:online}. Moreover, it might take six to eight times the length of the video for an untrained captionist \cite{ClosedCa43:online}. Another approach involved paying a professional company for captioning services, which might cost over \$8 per minute \cite{ClosedCa43:online}. 

\black{As automatic speech recognition (ASR) algorithms have advanced, some video producers have started using ASR to recognize and transcribe spoken language into readable text \cite{Useautom75:online,wactlar1996intelligent}. 
For example, some auto-captioning tools allow individual content creators to generate captions for their uploaded videos without human intervention \cite{fichten2014digital,Useautom75:online}. 
However, the variability and complexity of speech often cause issues regarding recognition accuracy, caption latency, and context formalization \cite{araujo2014context,gaur2016effects,kafle2016effect,kafle2019predicting,kushalnagar2014accessibility,o2008automatic,shiver2015evaluating}. 
Furthermore, background noise, multi-talker speech, human accent, and disfluent speech may further downgrade the quality of automatic captions \cite{benzeghiba2007automatic,glasser2017deaf,o2008automatic}.}
To make automatic captions work better, prior work explored the approaches such as removing the noise from the environment and changing the appearances of the automatic captions to convey ASR confidence \cite{berke2017deaf} (e.g., alternating the font size \cite{piquard2015qualitative}, font color \cite{shiver2015evaluating}, and underlining \cite{vertanen2008benefits}). Although these features were studied with DHH participants, they have not been fully integrated into video-sharing platforms, so it is unknown what the platform users will think of them in a practical scenario. Therefore, in our work, in addition to echoing some of the findings from the prior works, we reported the practices and perceptions of changing caption appearances on YouTube from the perspective of both individual content creators and DHH audiences.

Online video-sharing platforms, such as YouTube, enabled community contributions that allow video audiences or subscribers to contribute their effort to help caption YouTube videos \cite{Turnonma28:online}. Beyond the ASR approach, researchers explored the potential of leveraging community services or crowd workers for captioning \cite{huang2017leveraging,kushalnagar2012readability,lasecki2012online,lasecki2012real,lasecki2017scribe}. Due to the high cost of professional captionists and low accuracy of ASR, Lasecki et al. \cite{lasecki2012real} introduced a new approach that allowed groups of non-expert captionists to collectively caption speech and an algorithm for merging partial captions in real-time. To further improve the quality of captions created by non-expert captionists, Lasecki and Bigham \cite{lasecki2012online} presented methods that leveraged captionists quality estimation and caption quality estimation to overlap between non-expert captionists to provide optimal caption output. Moreover, Lasecki et al. \cite{lasecki2017scribe} further support non-expert captionists by managing the task load, directing different captionists to different portions of the audio stream, and adaptively determining the segment length based on each individual's typing speed.
Furthermore, Huang et al. \cite{huang2017leveraging} implemented BandCaption, a system that combines the ASR with crowd input to provide a cost-efficient captioning approach for online videos and distributed micro-tasks to crowd workers who have different strengths and needs. Huang et al. \cite{huang2017leveraging} conducted studies with people with different backgrounds and showed that different user groups could make complementary contributions based on their strengths and constraints. Although past research explored how to better support community captioning with non-expert captionists and evaluated their methods through user studies, it is unknown: 1) how community captioning was adopted by video-sharing platforms? 2) what are the practices and challenges for individual content creators to leverage community captioning to create the video captions? 3) what is the general quality of captions generated through community captioning in video-sharing platforms? 4) what are the concerns of community captions from the perspective of DHH audiences?

\red{Even though prior research has explored DHH audiences' perceptions on caption qualities of videos in general \cite{shiver2015evaluating}, there lack explorations of DHH audiences' experiences and challenges with captions of videos uploaded by individual content creators, who have special situations and background which may affect caption qualities. On the other hand, little research has widely explored the current practical approaches and challenges for individual content creators to caption their online videos on video-sharing platforms and associated challenges. In our work, we first present the feedback towards and potential improvement of the captions provided by individual content creators from DHH audiences' perspective through semi-structured interviews. Through video analysis and online surveys with YouTube content creators, we demonstrate existing captioning methods and processes, caption quality and consequences, and individual content creators' perceptions and general challenges of captioning.}
  

\section{Method}
\black{We first conducted semi-structured interviews with 13 DHH YouTube viewers to explore RQ1. To understand RQ2 and RQ3, we then conducted a YouTube video analysis with 189 non-commercial YouTube videos uploaded by individual content creators and a follow-up survey with 62 individual content creators. The two studies enable us to explore online video captions from both perspectives of caption creators and DHH consumers. In this section, we describe the methodological details of our studies.}

\subsection{\black{Semi-structured Interview}}
\black{In this section, we first show the study procedure of our interviews with DHH audiences (e.g., recruitment, demographic information, sample interview questions) and then present how we analyzed the data from interviews.}

\begin{table}[ht]
\small
\caption{Interviewees' Demographic Information} 
\centering 

\begin{tabular}{|p{1.3cm}|p{0.4cm}|p{1cm}|p{5.5cm}|p{3cm}|} 

\hline 
Participant & Age & Gender & Hearing Impairment Condition & Frequency of Watching YouTube\\ [0.5ex] 
\hline 
P1 & 50 & Female & Congenitally deaf & Nearly everyday\\
\hline
P2 & 26 & Female & Bilateral sensorineural hearing loss in 2014 & Everyday\\
\hline
P3 & 22 & Male & Acquired hard of hearing since teenager & Everyday\\
\hline
P4 & 35 & Female & Acquired hard of hearing since teenager & Everyday\\
\hline
P5 & 36 & Male & Acquired hard of hearing 12 years ago & Nearly Everyday\\
\hline
P6 & 25 & Female & Congenitally deaf & At least once a week\\
\hline
P7 & 34 & Female & Congenitally deaf & Nearly everyday\\
\hline
P8 & 18 & Female & Congenitally deaf & Everyday\\
\hline
P9 & 32 & Female & Congenitally deaf & Everyday\\
\hline
P10 & 28 & Female & Congenitally deaf & Everyday\\
\hline
P11 & 26 & Female & Congenitally deaf & Everyday\\
\hline
P12 & 63 & Male & Congenitally deaf & Everyday\\
\hline
P13 & 50 & Female & Congenitally deaf & Everyday\\[0.5ex] 
\hline 
\end{tabular}

\label{table:demographic} 
\end{table}

\subsubsection{\black{Study Procedure}}
To understand the quality of captions created by individual content creators from the perspectives of DHH audiences, we conducted semi-structured interviews with 13 YouTube audiences who are Deaf or Hard-of-Hearing (Table \ref{table:demographic}). Our interviewees have an average age of 34, with a range from 18 to 63 years old. Nine of them are congenitally deaf, and four of them acquired deafness in their teenagers. Interviewees were recruited through social platforms (e.g., Reddit, Twitter, Facebook). To participate in our interview, the interviewee must 1) be an audience of YouTube videos; 2) be deaf or hard-of-hearing; 3) be 18 or above; 4) be able to read and write in English. The interviews were conducted through Zoom and took around 45 - 60 minutes for each interviewee. In Zoom interviews, the communication was either done through typing on Zoom chat or speech if the interviewee has an automatic speech recognition software on their computer. 

In the interview, we first asked DHH participants about their demographic information (e.g., age, gender). We then asked them about their experiences of watching online videos created by individual content creators on video-sharing platforms (e.g., searching, filtering, watching, and commenting) and their associated concerns and challenges \black{(e.g., ``have you had experiences of watching a YouTube video that did not have captions once it got uploaded? If yes, please talk about the details.'')}. Afterward, we asked them regarding their perceptions of video caption on YouTube and their reactions towards poor quality captions \black{(e.g., ``what actions would you take if you come across a YouTube video with missing or poor-quality captions?'')}. Finally, we asked them about their recommended improvements on captions to improve the experiences of watching videos created by individual content creators \black{(e.g., ``what elements do you wish to add to Youtube captions?'')}. Interviewees who completed the interview were compensated by \$15 cash. The whole recruitment and study procedure was approved by the institutional review board (IRB).

\subsubsection{\black{Data Analysis}}
\black{After finishing interviews with 13 DHH YouTube viewers, we first combined all transcripts from different interviewees in the same folder. Two researchers then downloaded the folder in their local drive and independently performed open-coding on the transcripts. The researchers coded the transcripts focused on DHH audiences' practices, perceptions, and challenges on videos created by individual content creators. During the coding process, both researchers went through the transcripts multiple times.} Then, the coders met and discussed their codes. When there was a conflict, they explained their rationale for their code to each other and discussed to resolve the conflict. Eventually, they reached a consensus and consolidated the list of codes. Afterward, they performed affinity diagramming \cite{hartson2012ux} \black{through a Miro board \cite{AnOnline70:online}} to group the codes and identified the themes emerging from the groups of codes. Overall, we established three themes and 11 codes. The results introduced in the findings are organized based on these themes and codes. 

\subsection{\black{YouTube Video Analysis + Survey}}
\black{To further understand individual content creators' practices, challenges, and perceptions on captioning their own videos, we conducted a mixed-methods study by including two phases: 1) a YouTube video analysis---searching, filtering, and analyzing YouTube videos related to captioning practices and challenges from the perspective of individual content creators and 2) an online survey with YouTube content creators who captioned their videos---seeking an in-depth understanding of existing practices and challenges to caption videos from the perspective of individual content creators.}


\begin{table}[ht]
\caption{Searching Keywords} 
\centering 
\begin{tabular}{|p{8cm}|} 

\hline 
\textbf{Searching Keywords} \\ 
\hline 
Caption, Craption, Community Caption, Closed Captioning, Captioning, YouTube Caption, YouTube Automatic Caption, Automatic Caption, Self-captioning, Contribute Closed Caption, Content Creator Captioning, Captioning Services, Caption Accessibility, Caption Challenge, Caption Video, Video Accessibility, Captionist, Community Captionist \\
\hline 
\textbf{Hashtag Searching Keywords} \\
\hline
\#NoMoreCraptions, \#CaptionYourVideos, \#WithCaptions, \#ClosedCaptions, \#CaptionThis, \#CaptionPlease, \#CaptionVideos, \#AutomaticCaption, \#SaveCommunityCaption, \#ContributeClosedCaptions \\
\hline
\end{tabular}
\label{table:searchterms} 
\end{table}

\subsubsection{\black{YouTube Video Analysis---Data Collection}}
\black{Inspired by prior research on leveraging the richness of YouTube video contents to understand accessibility needs \cite{anthony2013analyzing,li2021non}, we conducted a YouTube video analysis to understand captioning practices and challenges from the perspective of individual content creators and find the potential reasons for caption problems that were complained about by our DHH interviewees from the previous section.}
 

\black{In the video analysis, we look for videos focused on captioning practices, challenges, and perceptions from individual content creators to uncover potential design implications and knowledge gaps, for example, individual content creators explaining their captioning methods and associated concerns. To search for the relevant videos about captioning, three researchers independently combined searching captioning keywords (e.g., YouTube captions, automatic captions, community captions) and leveraging the hashtag searching method \cite{gao2017hashtag,yang2012we} related to captions (e.g., \#NoMoreCraptions, \#CaptionYourVideos) (Table \ref{table:searchterms}). To come up with these searching keywords, our researchers first started with basic searching keywords (e.g., Caption, YouTube Caption) and gradually included other searching keyword combinations and hashtag keywords found from candidate video titles or descriptions. For example, we found the hashtag searching term `\#CaptionYourVideos' listed in the title from a video we searched by using ``YouTube captions''. Because each search may generate hundreds of results, we then followed the same approach as Komkaite et al. \cite{komkaite2019underneath} by stopping searching for videos after the whole page of searching results started to be irrelevant.}

In total, we initially created a video dataset of 248 relevant videos found by July 10th, 2020. We then filtered out videos if they: 1) were commercial videos that were not self-funded by content creators; 2) presented in a non-English language and did not have English caption; 3) were irrelevant to captions or videos created by individual content creators, such as captions created by professional captionists for movies companies; 4) were duplicated. We then ended up filtering 59 videos and created the final video dataset with 189 videos (V1 - V189). Among 189 videos in our dataset, most videos were uploaded in 2016 (68), while others were uploaded in 2019 (31), 2017 (24), 2020 (23), 2018 (18). The average length of videos was 354 seconds (ranging from 27 seconds to 1157 seconds).

\subsubsection{\black{YouTube Video Analysis---Data Analysis}}
\black{To code the videos, three researchers first open coded the videos independently. Then, the coders met and discussed their codes. When there was a conflict, they explained their rationale for their code to each other and discussed to resolve the conflict. Eventually, they reached a consensus and consolidated the list of codes. Afterward, we performed affinity diagramming \cite{hartson2012ux} to group the codes and identified the themes emerging from the groups of codes. Overall, we established three themes and 11 codes. The results introduced in the findings are organized based on these themes and codes.}

\subsubsection{Survey with YouTube Content Creators}

\begin{table}[ht]
\small
\caption{Survey Respondents Demographic Information} 
\centering 
\begin{tabular}{|p{1.4cm}|c|c|} 

\hline 
Category & Detail & Count\\ [0.5ex] 
\hline 
\multirow{6}{4em}{Age} & 18 - 24 & 22 \\
& 25 - 34 & 23 \\
& 35 - 44 & 8 \\
& 45 - 54 & 4 \\
& 55 - 64 & 3 \\
& 65 or above & 2 \\
\hline 
\multirow{5}{4em}{Gender} & Male & 32 \\
& Female & 23 \\
& Non-binary & 4 \\
& Prefer not to disclose & 1 \\
& Prefer to self-describe & 2 \\
\hline
\multirow{10}{4em}{Primary Occupation} & Management & 1 \\
& Business and Financial Operation & 4 \\
& Computer and Mathematical & 6 \\
& Architecture and Engineering & 1 \\
& Life, Physical, and Social Science & 2 \\
& Educational Instruction and Library & 20 \\
& Arts, Design, Entertainment, Sports, and Media & 15 \\
& Healthcare Practitioners and Technical & 1 \\
& Food Preparation and Serving Related & 1 \\
& Sales and Related & 5 \\
& Unemployed or Retired & 5 \\
\hline
\multirow{6}{4em}{Length as a Content Creator (Years)} & Less than One Year & 9 \\
& One to Two Years & 18 \\
& Two to Three Years & 8 \\
& Three to Four Years & 5 \\
& Four to Five Years & 8 \\
& Above Five Years & 14 \\

\hline
\multirow{5}{4em}{Subscribers} & Over 1 million & 0 \\
& 100k - 1 million & 5 \\
& 10k - 100k & 9 \\
& 1k - 10k & 14 \\
& less than 1k & 34 \\
\hline

\end{tabular}
\label{table:surveydemographic} 
\end{table}

Inspired by previous research \cite{anthony2013analyzing,komkaite2019underneath}, we further conducted an online survey study with YouTube content creators to acquire more details regarding content creators' practices and challenges through captioning and to address uncertainties from the video analysis. To find content creators who have experience captioned videos before, we first contacted YouTube content creators of the YouTube videos we collected and posted our recruitment script in the YouTube comment section on the videos we collected. Due to the limited response rate, we further distributed our recruitment script through social platforms (e.g., Reddit, Twitter). To participate in our online survey study, the participant must 1) be an individual content creator on YouTube; 2) have experience creating captions on her uploaded videos; 3) be 18 or above; 4) be able to read and write in English. The survey was hosted through Qualtrics \cite{Qualtric28:online}. The survey questionnaire included 22 questions covering demographics (e.g., age, gender, upload frequency, primary occupation), practices of captioning (e.g., what captioning methods have you used the most? Why?), perceptions on captioning, and challenges associated with captioning as an individual content creator (e.g., Have you had any difficulties when using community-contributed captions? Why?). \red{To encourage authentic reporting and protect participants’ privacy, we asked participants to select the number of subscribers under different intervals instead of providing the exact number of subscribers (i.e., less than 1k, 1k - 10k, 10k - 100k, 100k - 1 million, over 1 million) (Table \ref{table:surveydemographic}) \cite{YouTubeC49:online}}. Survey respondents who completed the online survey were entered into a draw for a \$100 Amazon gift card. In total, we received 62 responses from YouTube content creators (S1 - S62) (Table \ref{table:surveydemographic}). The whole recruitment and study procedure was approved by the institutional review board (IRB).

\section{Results}
\black{In the results, we will first present the findings from the interview with DHH audiences on RQ1. We will then demonstrate the findings from the YouTube video analysis and survey with individual content creators to answer RQ2 and RQ3.}

\subsection{\black{DHH Audiences' Practices, Perceptions, and Challenges on Videos Created by Individual Content Creators}}
In this section, we show our findings in three main phases regarding DHH audiences' experiences with videos created by individual content creators: 1) video searching and filtering, 2) feedback on caption quality and improvements, and 3) DHH audiences' reactions and mitigation strategies.

\subsubsection{Video Searching and Filtering}
\label{DHH practice}
\begin{figure}
    \centering
    \includegraphics[width=0.8\columnwidth]{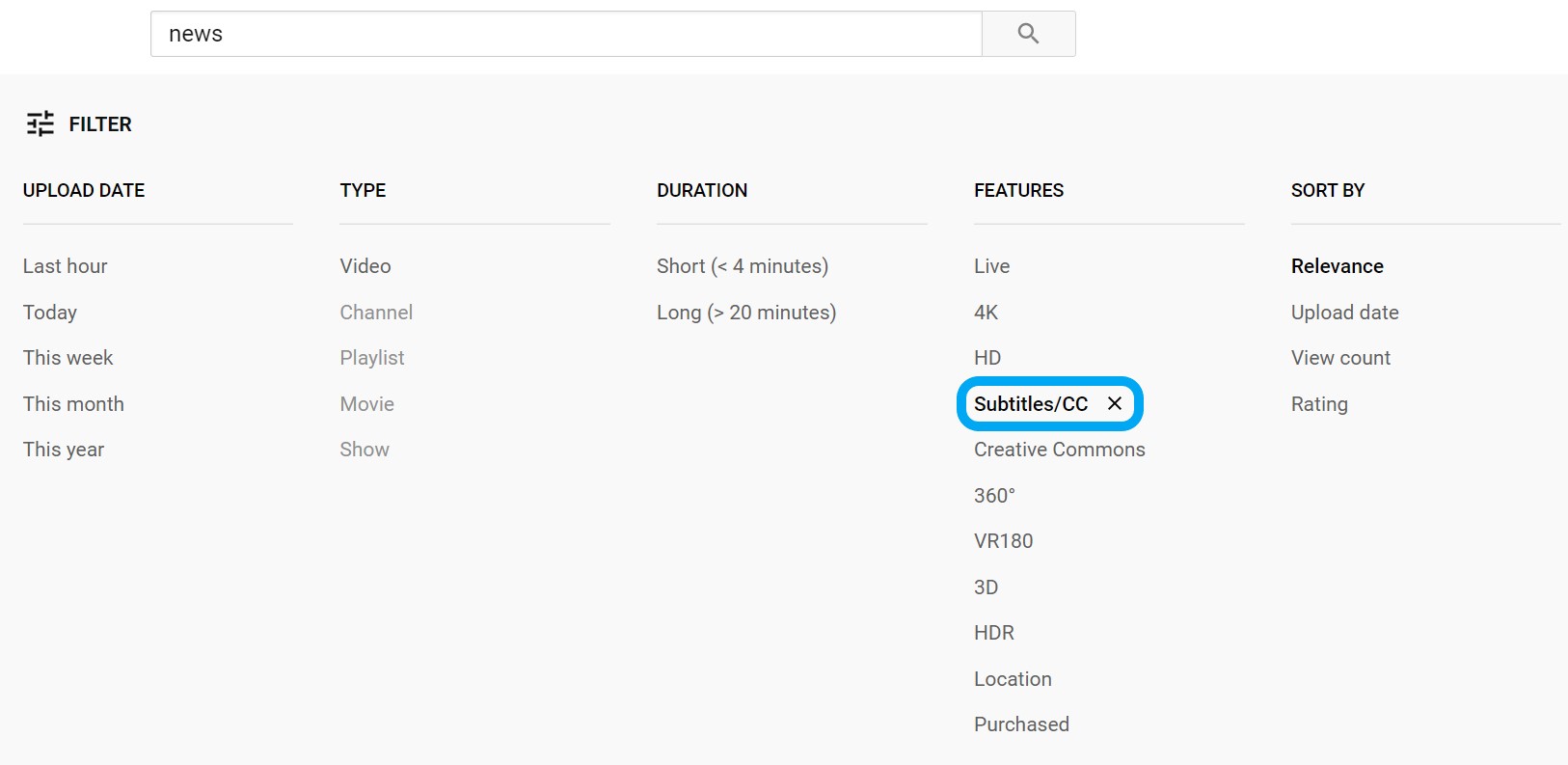}
    \caption{YouTube video filter in searching. The system will only display videos with captions if the audience selects the subtitle/CC filter.}
    \Description{There is a searching bar at the top of the figure. There are five main categories of filters in the figure, which includes upload date, type, duration, features, and sort by. There is a blue highlight that circles Subtitles/CC under features.}
    \label{fig:youtubefilter}
\end{figure}

In our interview, we asked our participants to describe their experiences and practices of searching and filtering their videos of interest in video sharing platforms. We found that they commonly used two approaches to find videos of interests: 1) using the searching filter to only display videos with captions (Fig. \ref{fig:youtubefilter}) and 2) directly exploring videos from the feed list that are automatically recommended by video-sharing platforms. Although having filters on videos with/out captions reduces the searching effort for DHH audiences, seven participants mentioned that the caption quality varied on different captioned videos. Because the filter is based on whether the video has captions or not, not the caption quality. Therefore, we found that DHH audiences tend to watch videos that content creators manually added `[CC]', `\#Captioned', or `\#WithCaption' either in the video title or in the video description. \black{\textit{``Having manual tags made me feel that the content creator has the awareness of providing captions, and it makes me trust the content is accessible.''} said P8.} Although these \textbf{manually added caption tags could create additional confidence and trust in caption qualities for DHH audiences}, P5 commented on the lack of manually added caption tags and the variation of styles, which made him hard to search:

\begin{quote}
    ``...The existing Subtitle or CC filter can only tell whether a video is captioned or not. From my own experiences, most of them are crappy. Afterward, I started searching for videos that the video creator manually put CC tags in the title or descriptions. This shows that the video creator considers video captions more seriously. However, I found that only less than 1\% of videos have manually added caption tags, and the styles of those tags really varied on different videos. For example, content creators added self-created captions tags like `[CC]', `cc', `[Captioned]', and `WithCaption'. These added tags gave me a hard time in video searching...''
\end{quote}

To reduce the effort DHH audiences spend manually passing through the videos and checking proper captions, DHH audiences also stated that video-sharing platforms should include the predictions of the caption quality of videos (P2, P13). \black{\textit{``Normally, If I want to make sure a video has proper captions, I have to physically open it and preview it for a while to know if the caption is readable. I do not want to waste my time on watching videos to just make sure it has proper captions.''} said P2.} As the quality measure metrics, prior research has explored Word Error Rate (WER), Term Frequency-Inverse Document Frequency (TF-IDF) \cite{nanjo2005new}, Match Error Rate (MER) \cite{morris2004and}, and Automated-Caption Evaluation (ACE) \cite{kafle2017evaluating}. However, participants reported that most video-sharing platforms \textbf{lack caption quality indications} for their videos. They stated that video-sharing platform designers should focus on choosing the most efficient quality measure metrics and explore how to visualize the caption quality indications. P13 explained her practices of checking caption qualities and needs of quality indications:

\begin{quote}
    ``...Every time when I find videos from the feed list, unless it mentioned [CC] in the title, I have to physically open it and check if it has proper captions. This requires me to open the video from the feed list, manually turn on the caption option, and watch the video for a while to know whether I could understand the video content from the captions. If YouTube or other platforms could show whether a video has captions and what is the confidence of the caption quality, it would save my time checking the captions, and I could even add filters to remove certain videos below the quality baseline...''
\end{quote}

In terms of directly exploring videos in the feed list, nine participants mentioned the delay of captions to be available on videos while they are already published and presented in the recommendation list. \black{According to the literature, manual captioning methods might take six to eight times the video's length to create the captions by unprofessional captionists \cite{ClosedCa43:online}. \textit{``Many videos were first uploaded and showed on the platform without any captions, because many content creators want their videos uploaded at a certain time to receive more views. When the captions were ready, content creators then added the captions to the videos.''} said P3. The \textbf{caption delay forced DHH audiences to manually save videos of interest for the future} while searching videos. Otherwise, it disappears in the feed list of video-sharing platforms.}
P12 emphasized why delay is the biggest problem:

\begin{quote}
    ``...There are two key problems with the caption delay. First, it may disappear from the feed list of the audiences when it gets the caption ready because the feed list usually displays the most recent videos. Second, I often lose patience while waiting for captions...''
\end{quote}

\subsubsection{Feedback on Caption Quality and Improvements}
\label{feedback caption quality}
\begin{figure}
    \centering
    \includegraphics[width=0.6\columnwidth]{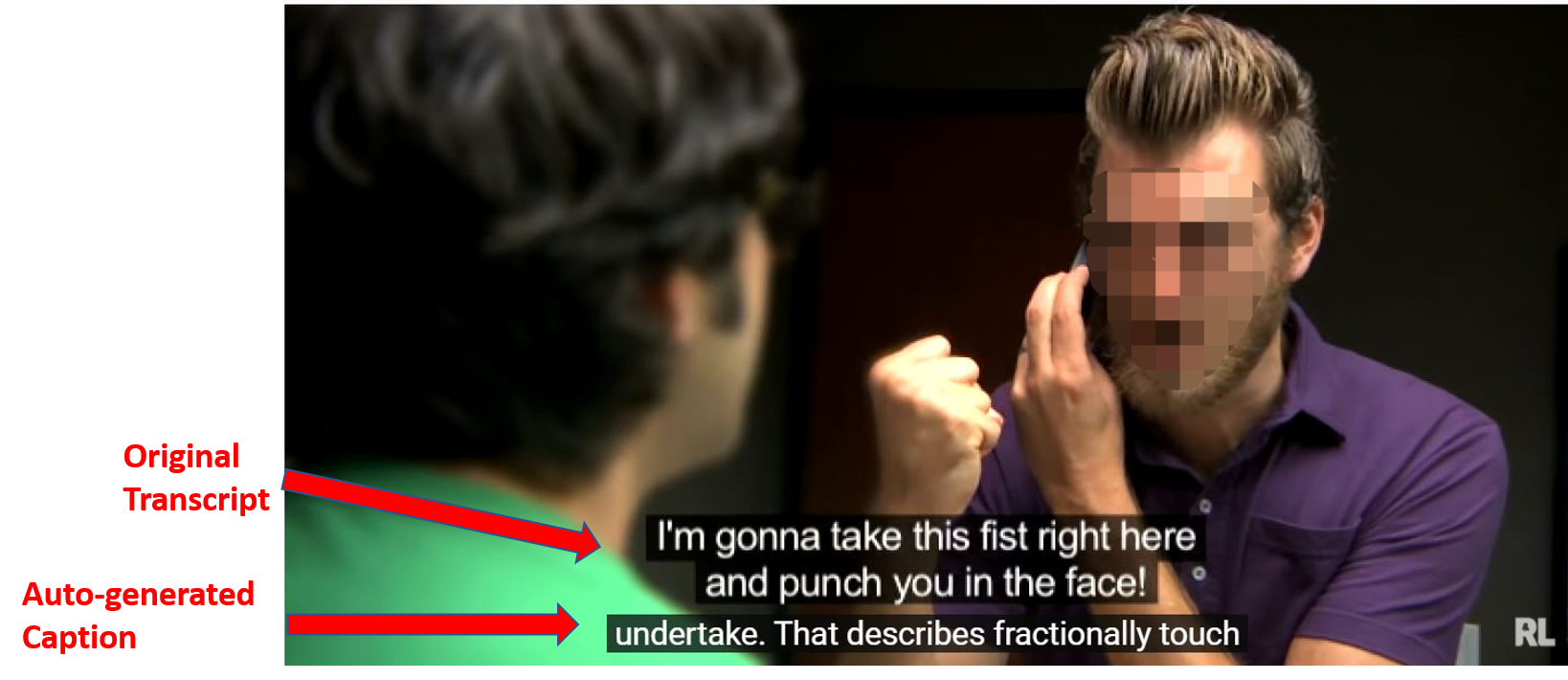}
    \caption{This figure compares the auto-generated caption (the bottom line) with the original transcript (the top two lines).}
    \Description{This figure includes two people. There is an original transcript ``I'm gonna take this fist right here and punch you in the face!'' and there is an auto-generated caption as ``undertake. Take describes fractionally touch''}
    \label{fig:autocaptionproblem}
\end{figure}

We further asked our participants regarding their perspectives of video caption qualities on YouTube and potential improvements. In this section, we uncover DHH audiences' feedback on caption styles and presentations, appearances control, and other peripheral visual contents, which may affect the experience with captions. All of our interviewees complained that the majority of videos created by individual content creators did not even have a caption or had poor-quality captions that were useless for understanding the content. 
More specifically, We found that spelling, grammar, and punctuation problems are present in both auto-generated captions (e.g., Fig. \ref{fig:autocaptionproblem}) and human-generated captions. Among the many problems (e.g., spelling, punctuation, grammar), eight of our participants emphasized that \textbf{punctuation is more critical to understanding the content of the caption over other problems.} \black{They mentioned that errors such as spelling and grammar would not affect the understanding of the captions too much because they could predict the correct one from the caption context. However, punctuation problems may mess up the comprehension of the captions. For example, one sentence may not fit in one single caption and the second half of the first sentence might be displayed together with the first half of the second sentence (P4, P6, P7).} P4 stated the importance of punctuation:

\begin{quote}
    ``...Spelling is not that important to me, although I do prefer perfect spelling. I could sort of get the correct word from the content background and my personal experiences on predictions. But punctuation is definitely more important! If I could know where a stop is, it could help me to understand the content of the sentence. Without punctuation, it puts lots of mental effort while watching the videos on YouTube...''
\end{quote}

Beyond common caption errors, four participants reported that some captions tend to use `[]' to abbreviate certain video content. We found this \textbf{use of abbreviation in videos generates concerns on equality and fairness for the DHH population}.
For example, P2, P5, and P6 mentioned their experiences of watching some video captions that put `[Joke]' instead of the actual joke content, which made DHH audiences frustrated when watching the video with their friends and family members. P2 commented:

\begin{quote}
    ``...I was really frustrated when the caption just put [Joke] when they are actually telling a joke in the audio. It just makes the whole joke very very flat. Especially, it felt embarrassing when I watched a video with my friends, and they all laughed, and I just did not know what happened...''
\end{quote}

Moreover, our participants mentioned that some captions used ``[Music]'' in the caption to represent musical content. To better describe audio content in captions, all of our DHH interviewees suggested adding more detailed information that helps with identifying the content, such as ``[Rap Music done by XXX]'' rather than just ``[Music].'' Beyond that, DHH audiences also want the caption to include visually unidentifiable or hard-to-identify audio information in the caption, such as ``cough'', ``laugh'', and ``inhale''. On the other hand, three participants also posed concerns about over-detailed descriptions, which may distract DHH audiences. P9 explained the importance of \textbf{having a certain hierarchy in captions and allowing audiences to control the level of details}:

\begin{quote}
    ``...Honestly, I do prefer having detailed captions that describe things that are relevant to the videos. However, I have watched a movie where they added background music lyrics and all character identities in the caption and that really freaked me out. I think there should be a hierarchy in captions based on the importance to understand the video content. And I should be able to control the level of details in captions...''
\end{quote}

In terms of caption presentations, our interviewees also complained about existing presentation styles, such as font size \cite{piquard2015qualitative}, font color \cite{shiver2015evaluating}, and underlining \cite{vertanen2008benefits}, which is pre-defined by individual content creators. Five interviewees expressed \textbf{strong preferences for controlling the changes of caption appearances and the correlation between the changes and the meaning of the changes}. For example, P12 mentioned his willingness to change the font sizes according to different devices because he is 63 and often has a hard time with different caption font sizes on his phone, tablet, and laptop.

\begin{figure}
    \centering
    \includegraphics[width=0.5\columnwidth]{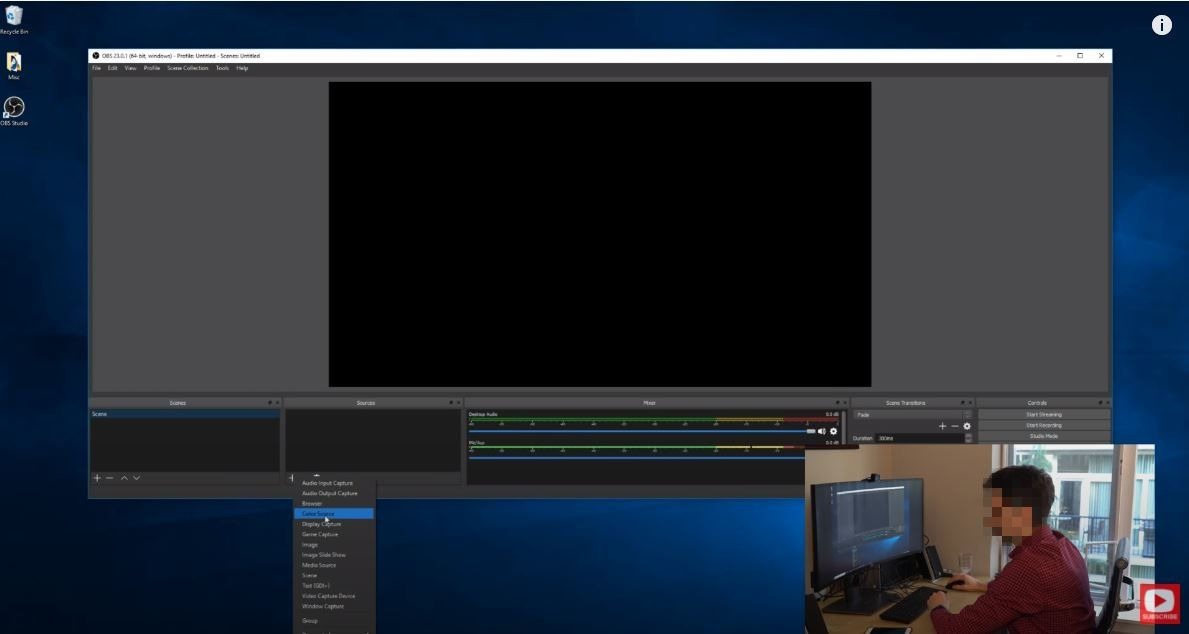}
    \caption{This figure shows the streamer explaining how to use the recording function on Windows, with a small head at the bottom right corner. The streamer's face does not directly face the camera, which makes DHH audiences hard to read his lips.}
    \Description{This figure shows the streamer explaining how to use the recording function on Windows, with a small head at the bottom right corner. The streamer's face does not directly face the camera, which makes DHH audiences hard to read his lips.}
    \label{fig:lipreading}
\end{figure}

Captions are not the only visual content that helps DHH audiences understand the video content. In our interview, our participants stated the importance of lip-reading to understand the video content. More importantly, our participants complained that some videos on online video platforms only show a \textbf{small talking head at the corners of the screen that made it difficult for viewers to read the speaker's lips}. Moreover, some do not directly face the camera, which makes lip-reading impossible (Fig. \ref{fig:lipreading}). P3 stated the importance of reading lips when watching a video on YouTube:

\begin{quote}
    ``...I would say I highly rely on lip-reading when watching YouTube videos. It is more peripheral if the caption quality is good, but it is definitely more helpful if the video has poor-quality captions. In some videos, the content creators directly face the camera, which is fine for me to read lips. But, many videos, such as game streaming videos where the content creators usually show their minimized face at a corner of the screen really pissed me off. Some of them even do not show their faces or at a weird camera angle that I cannot even see the mouth. I would recommend the social media platform could provide an option to maximize the lip region and display it on the screen...''
\end{quote}

\subsubsection{DHH Audiences' Reactions and Mitigation Strategies}

To check the availability of video captions due to caption delay, nine participants complained about the tedious process of manually going through the video list. Thus, six out of 13 interviewees mentioned that they \textbf{typically do not come back to a video for captions if the video was not properly captioned initially when they first saw the video in the feed list}. They would watch a different video that has similar content instead. Therefore, this posed concerns to individual content creators to create captions on time to prevent the loss of audiences.

In terms of reactions to caption problems, we found that 10 out of 13 DHH interviewees mentioned that they tried leaving a comment initially but ended up with other options (e.g., emailing, direct messaging on Instagram if applicable) or just did not do anything. \textbf{Leaving a caption request in the commenting section might not be an effective approach}, as P11 commented that her comments on requesting captions usually got buried in the long comment list, especially for big YouTubers:

\begin{quote}
    ``...Initially, when I found videos with bad captions on YouTube, I left comments to let the content creators know my concerns. However, I found the content creators barely check their comments, especially those with thousands of comments. Mine just got buried at the bottom...''
\end{quote}

P4 also mentioned that she resists posting comments to ask for captions because other video audiences piled on her by replying disrespectfully, such as ``you are wasting the content creator's time.'' Therefore, our interviewees mentioned that they tend to contact content creators more privately by trying other direct messaging approaches, such as Twitter and Instagram, to request better captions. \black{However, three interviewees further mentioned that the additional effort of sending private messages (e.g., sign up a new account and send friend requests) to individual content creators to request better captions and potential privacy concerns made them end up choosing a different video. Overall, all participants mentioned that \textbf{the process of requesting better captions is troublesome and useless} and future research should explore new channels of communication about captioning problems with privacy protections. P4 commented on the burden of creating additional accounts for private messages and the timely responses:}

\begin{quote}
    ``...Different YouTubers have their own preferred platforms for personal or business inquiries. If I do not have an account with a certain platform, I have to sign up a new account and send a friend request to the YouTuber in order to send private messages to the person. Honestly, it may still take a long time to have the YouTuber replied to my message or just never replied...''
\end{quote}

In this section, we first revealed DHH audiences' experiences on video searching and filtering (e.g., manually added caption tags could create additional honest signals and trust to DHH audiences). We then uncovered DHH audiences' feedback on caption quality by individual content creators (e.g., use of abbreviation in videos generate concerns on equality and fairness for the DHH population). Finally, we showed their reactions and mitigation strategies on low-quality captions and associated concerns (e.g., attacked with disrespectful replies by other video viewers). From the interview, we explored existing practices, challenges, and perceptions on video captions from the perspectives of DHH audiences. To better understand the gap between DHH audiences and individual content creators, it is also important to explore individual content creators' practices, perceptions, and challenges in captioning their own videos.

\subsection{Individual Content Creators' Practices, Perceptions, and Challenges in Captioning Their Own Videos}
In this section, we show our findings in three main phases regarding individual content creators' practices, perceptions, and challenges on captioning: 1) caption practices and processes, 2) caption quality and consequences, and 3) individual perceptions and personal challenges on captioning.

\subsubsection{Caption Practices and Processes}
\label{caption practice and processes}
In this section, we introduce the existing practices for individual content creators to caption their videos and emphasize the practical challenges associated with different captioning approaches and processes. From the video analysis and the survey, we found that individual content creators mainly caption their videos through 1) self-captioning: content creators manually create the SRT file with timestamps or use auto-sync functionality to generate associated timestamps, 2) community-contributed captioning: video audiences and subscribers volunteer and contribute in the captioning process, 3) auto-generated captioning: captions automatically generated through speech recognition algorithms with/out manual editing by content creators, and 4) third-party captioning services: third-party companies generate the SRT file with captions, and content creators upload them later. In our survey, we asked participants to rate their preference on different captioning methods on a 5-point Likert scale (5 as strongly agree with preferring the method, 1 as strongly disagree with preferring the method). We found that our survey respondents preferred mostly on self-captioning (Mean = 3.56, SD = 1.6), automatic captioning with manual editing (Mean = 3.5, SD = 1.47), and community-contributed captioning (Mean = 3.1, SD = 1.6). On the contrary, our survey respondents showed the least preferences on third-party captioning services (Mean = 2.02, SD = 1.33) and automatic captioning without editing (Mean = 1.92, SD = 1.24). We further show the practices and challenges of each captioning method. 

\begin{figure}
    \centering
    \includegraphics[width=0.8\columnwidth]{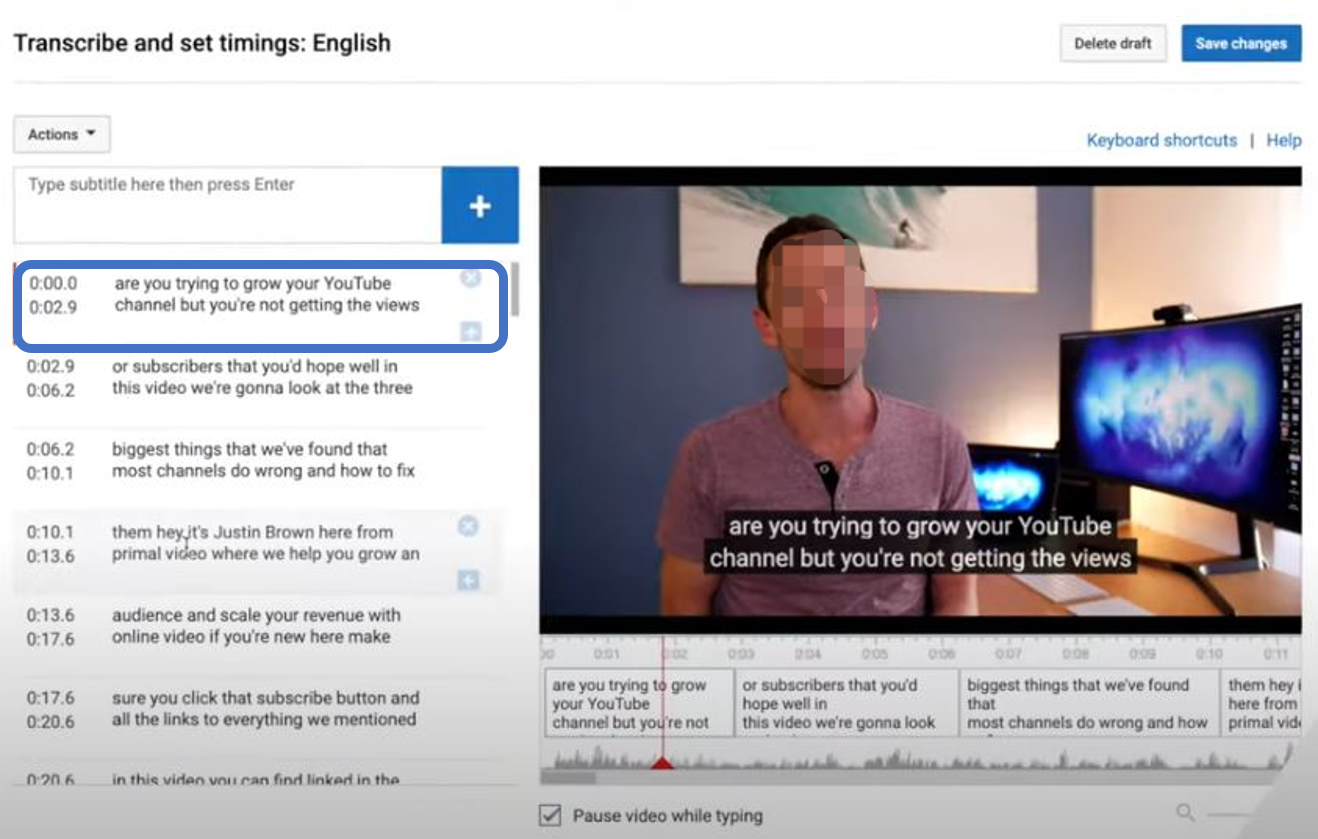}
    \caption{YouTube manual captioning interface and content creators could manually input the transcript in the automatically created time periods.}
    \Description{This shows a YouTube manual captioning interface. On the left side, there is a highlight that select a specific caption, there is a video frame shows at the right side with time stamp.}
    \label{fig:manualcaptioning}
\end{figure}

To caption videos manually, content creators caption their videos by uploading an SRT file or using YouTube captioning interface to input captions (Fig. \ref{fig:manualcaptioning}). In Fig. \ref{fig:manualcaptioning}, content creators could type their subtitles in the automatically generated time periods on YouTube. However, unlike professional video producers, individual content creators mentioned \textbf{the extra effort of reproducing video transcripts for self-captioning} because they often do not prepare a transcript before filming the video. Instead, they usually just have a general guideline or a checklist when filming videos (V25). In our survey, We found that 43 survey respondents (i.e., individual content creators), who captioned their videos by themselves, usually had to transcribe the entire video after filming it. Furthermore, 62 individual content creators mostly agreed that `manually captioning videos is time-demanding' (Mean = 3.92, SD = 1.27, on a 5-point Likert scale). We further asked individual content creators about their agreements with whether different steps of self-captioning are challenging on a 5-point Likert scale, 5 as strongly agree, and 1 as strongly disagree. We found that the top three challenging steps for self-captioning are typing the caption word by word (Mean = 3.86, SD = 1.32), syncing/tracing the timeline of captions (Mean = 3.72, SD = 1.32), and ensuring the captions are free of errors (Mean = 3.52, SD = 1.34).

\begin{figure}
    \centering
    \includegraphics[width=0.8\columnwidth]{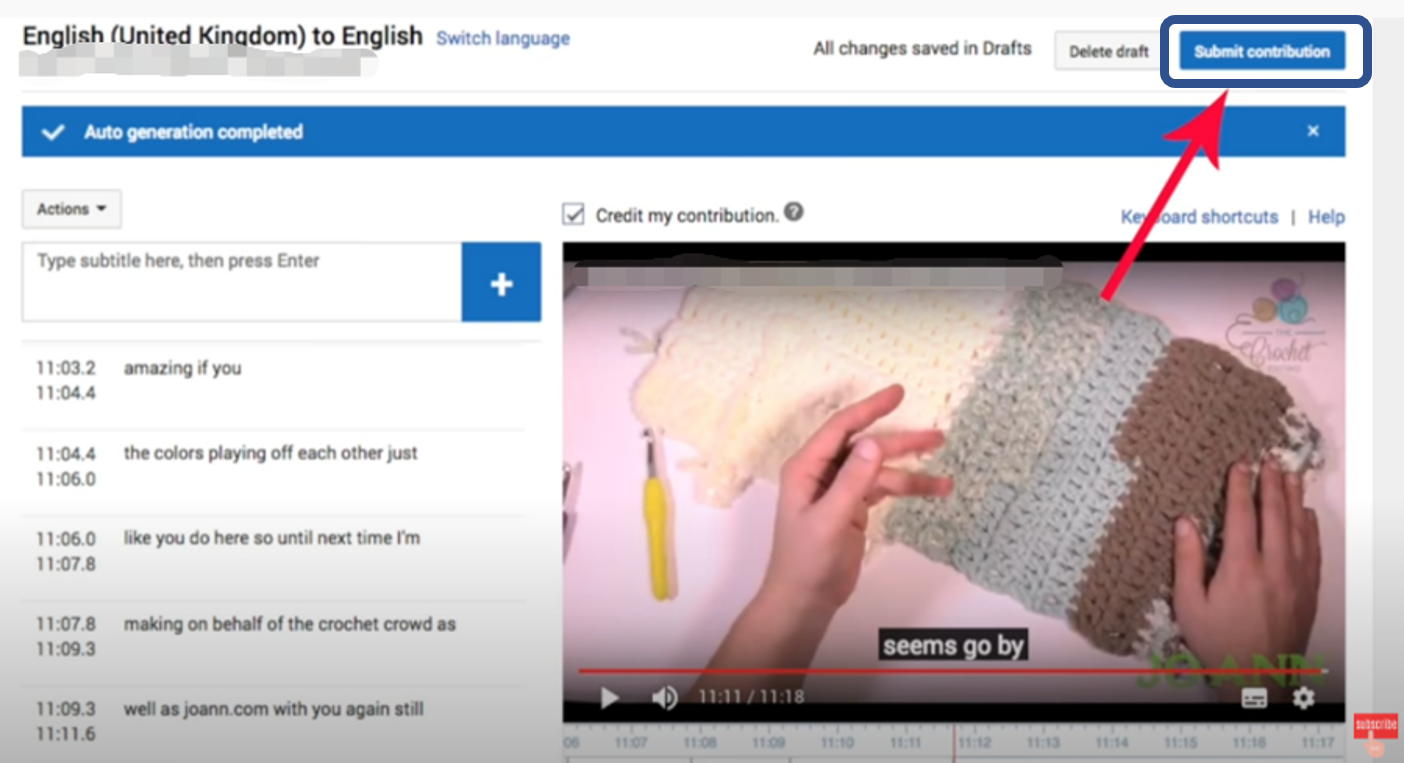}
    \caption{Community caption interface. Community captionists could manually input the transcript and submit it for approval by the individual content creator.}
    \Description{This figure shows a community caption interface with different captions with time stamps at the right. There is a video on the right side. At the top right, there is a button got highlighted as ``Submit contribution''}
    \label{fig:communitycaptioninterface}
\end{figure}

The community caption services in YouTube leverage audiences and subscribers to volunteer and contribute to the captioning process (e.g., V35) and also help to translate to other languages (V29). To get the video captioned through community contributions, content creators usually have to make the video public first. Then people who would like to contribute could add captions later. The community captioning interface (Fig. \ref{fig:communitycaptioninterface}) is similar to the self-captioning interface (Fig. \ref{fig:manualcaptioning}), except it has the button to submit the contribution for the additional approval process by the video owner. From our survey results, we found 46.8\% of our respondents leverage community caption services for captioning. To request community captions, we found that individual content creators chose to either post an individual video (V83), leave the request in the video description, or pin a comment in the video commenting section (V47). In our survey, we asked individual content creators to rate their agreement on practices of different approaches to request community-contributed captions on a 5-point Likert scale, 5 as strongly agree, and 1 as strongly disagree. We found the two of the most preferred approaches are leaving a notice in the video description (Mean = 3.28, SD = 1.39) or in the comments (Mean = 2.98, SD = 1.34). 

For the 53.2\% of our survey respondents who do not use community captioning or stopped using it, 78.1\% of them complained about the difficulty of finding enough volunteers and the long waiting time for the caption to get ready. From the video analysis and survey, we found that \textbf{individual content creators with fewer subscribers tend to wait longer to have captions ready through community-contributed captioning} (V140, V183). From our survey, we asked individual content creators on \textit{``how long does it usually take for community-contributed captions to be done?''}. Individual content creators, who used community-contributed captions, mentioned that it takes 16 - 24 hours on average to get their videos captioned through community captions. If we analyze the time cost to get the video captioned in two groups: subscribers higher than 1k and subscribers lower than 1k, we found that content creators with more than 1k subscribers only need to wait 8 - 16 hours on average, and it usually takes 1 to 3 days for content creators with less than 1k subscribers. In V140, the content creator commented:

\begin{quote}
    ``...I am not a big YouTuber who has millions of subscribers to help with captions. For small YouTubers like me, there are very few audiences who help me to caption my videos, and it usually takes a long time for my videos to get captioned...''
\end{quote}

Because community captionists are usually the video audiences of a specific channel, they have certain understandings of the video taste from the channel and are more willing to help video captions from the same channel. Due to the community caption delay, we found that a single individual content creator often has multiple videos from multiple channels that are pending for captions at the same time (V45). Therefore, they prefer directing community captionists from videos to videos for balancing the pending time of community captions (V90). However, individual content creators found that these \textbf{community captionists are unwilling to help captions for videos that they are not familiar with} which leads to even longer wait times for certain videos from different channels (V185). Due to the `anonymity' of community captioning, content creators often do not have much background understandings of these community captionists' preferences on videos. Therefore, individual content creators mentioned that different videos from different channels from the same content creator might have drastic time cost differences on captions through community captioning (V185).

In the previous section, we have shown that using auto-generated transcripts and modifying them to reduce speech recognition errors is the second most preferable method from our survey responses. In terms of the time cost, our survey respondents mentioned that they usually have to wait an average of four to eight hours for the automatic caption to be ready on YouTube. However, we found that the automatic captioning process of many video-sharing platforms is not a service that content creators could manage or control; the \textbf{auto-captions often appear automatically in the video once ASR is done without any notification} (S41). Therefore, we revealed that content creators, who have very limited time and rely on modifying automatic captions, recommended having video-sharing platforms send a notification message once the automatic captioning was finished, so they could check the caption quality immediately and prevent their audiences from reading error-prone captions generated automatically.

To save time and effort in captioning videos, content creators could pay third-party captioning companies for caption services. On average, it may cost from \$1 to \$5 per minute (V4). Comparing with the prior report in 2014 \cite{ClosedCa43:online}, we found that the cost for third-party captioning companies has decreased over the years, and some content creators with disabilities could usually negotiate for a discount (V3). Nevertheless, 43.5\% of survey respondents still mentioned that they do not have enough budget for third-party captioning services. We found that different approaches have their strengths and weaknesses, and there does not exist a single option that is a clear ``win'' for all groups of individual content creators. As a common practice, we found that \textbf{content creators usually stick to a single captioning approach without changing} (V82).

\subsubsection{Caption Quality and Consequences}

During our video analysis and survey, we found that content creators expressed their concerns on caption quality from auto-generated captions, community-contributed captions, and captions created by a third party. First, we found that \textbf{caption errors could potentially demonetize the video} (i.e., mark the video as not ad friendly) by falsely including non-ad-friendly words (e.g., swear words, racist words). From the video analysis, we found non-ad-friendly videos usually result in a lower placement in YouTube's recommendation system and eventually have fewer views (e.g., V171). In V59, the content creator explained his experiences and understanding of how poor captions containing racist words demonetized his video:

\begin{quote}
    ``...I was very surprised once I got my video yellow labeled (not ad friendly), which means I would basically not receive money from that video. I was pretty sure I did not say anything racist or contained any adult-only information that is not ad-friendly in the video. That video was captioned automatically, and I later found the caption contained racist words. Since then, I have stopped using auto-generated captions...''
\end{quote}

\begin{figure}
    \centering
    \includegraphics[width=0.5\columnwidth]{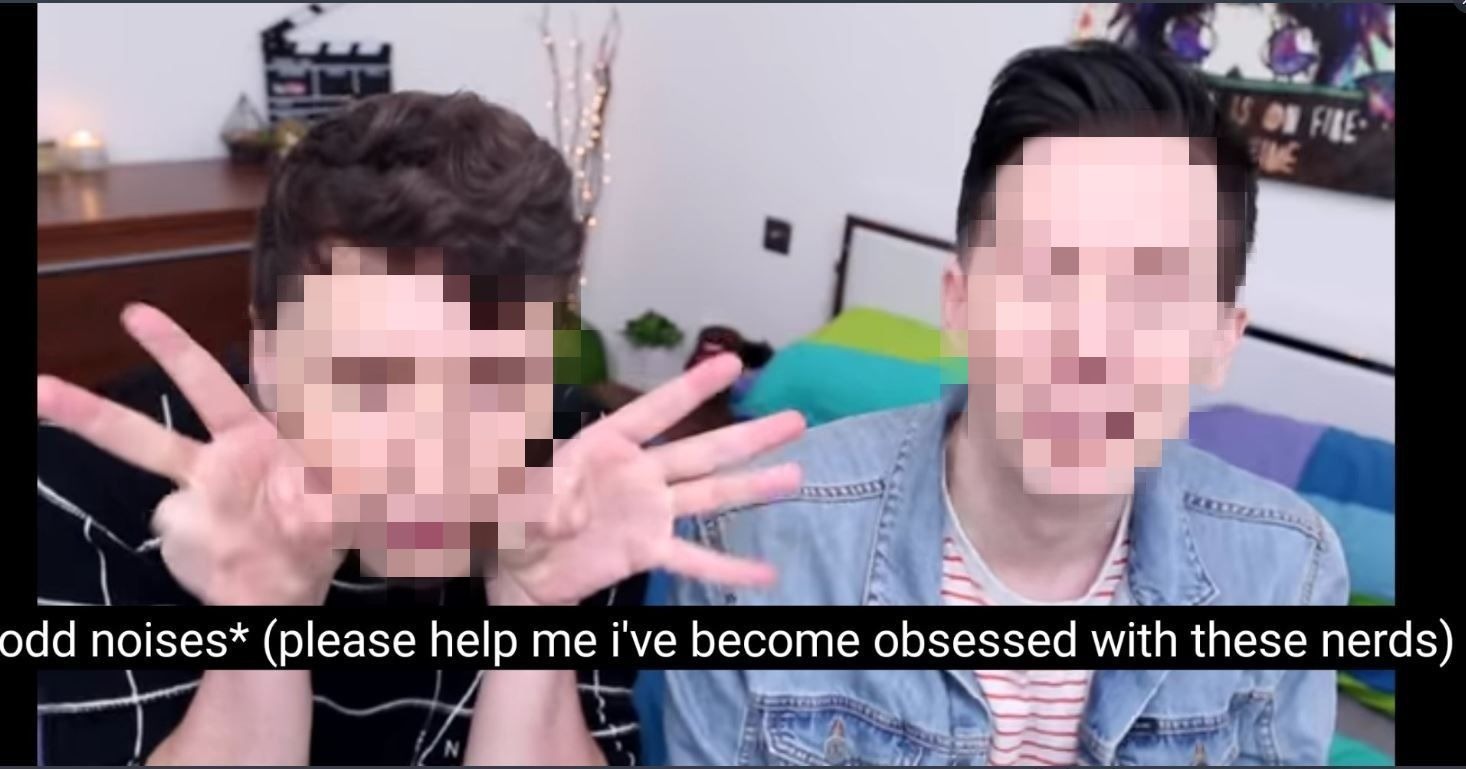}
    \caption{The caption includes personal comments from community captionists.}
    \Description{This figure has two people, one with black shirt and another one with a jacket. There is a caption in the video frame ``odd noises* (please help me i've become obsessed with these nerds)''}
    \label{fig:commentaryCaption}
\end{figure}

Second, 56\% of our survey respondents leverage community-contributed captions. However, some content creators complained about \textbf{unprofessional captionists sometimes adding their own comments or feelings in the community-contributed captions} (Fig. \ref{fig:commentaryCaption}). For example, promotion ads, irrelevant jokes, useless commentary, or racist speech (e.g., V58) could frustrate both content creators and their audiences. Although content creators need to approve the community captions in order for it to get published, we found that less than 1/7 of the 56\% survey respondents who use community-contributed captions spend additional time to verify the captions step by step. The majority of them directly approve and publish the captions when they are ready without checking the quality thoroughly due to limited time.

\subsubsection{Individual Perceptions and Personal Challenges on Captioning}
In this section, we report on content creators' perceptions, challenges, and awareness through captioning. From the video analysis and survey, we found that \textbf{content creators' individual circumstances limited their options of captioning methods}. Among the 62 content creators we surveyed, we found that only four of them consider being a content creator as their primary occupation, and many YouTubers either do not have enough money (e.g., V99) or enough time (e.g., V124) or even both to caption their videos. In V19, the content creator, who is a college student, explained her situation about captioning when auto-generated caption did not work for her because of her accent:

\begin{quote}
    ``...I know I should caption my videos, but I also have school deadlines, this becomes a dilemma for me. I am still on my student loan that I need to pay. Since my accent is pretty strong, it even took me more effort to modify the auto-generated captions than just manually create the transcript. I tried to post caption requests in my own channel, on Reddit, Facebook groups, but it barely helped...''
\end{quote}

Furthermore, we found \textbf{content creators with disabilities have strong accessibility needs on captioning and video editing tools}. From our survey, one survey respondent (S13) actually has hearing impairments and reported having a hard time using the current system to caption their videos, which forced her to rely on community captions or pay third-party companies for captions. S13 left a comment in the survey about her situation:

\begin{quote}
    ``...As a content creator with hearing impairments, I face strong difficulties through video editing processes. To caption a video, the only option for me right now is to pay third-party companies to do all the work for me. Although these captioning companies gave me the discount as \$1 per minute, it is still a considerable amount for my budget...''
\end{quote}

Beyond the physical inabilities, we found that the circumstances of having \textbf{poor-quality captions are also caused by the lack of awareness and understanding of captioning as an individual content creator}. In our survey, we asked content creators how they initially learned the importance and methods of captioning. We found that 43.5\% of the content creators said they learned by watching tutorial videos, and 24.2\% by reading articles about captions voluntarily. However, this voluntary learning requires content creators to have an awareness of captioning videos. Only 9.7\% of the content creators learned that they should caption their video through requests from viewers' comments or direct messages.

From the video analysis and survey, we found that many individual content creators are not aware of the importance of having captions when uploading videos to video-sharing problems. Later after content creators learned the importance of captioning videos, they often put themselves under pressure to back-caption all of their past videos. We learned that this large effort of \textbf{back-captioning (i.e., caption all previous videos that do not have captions) problems} made the whole captioning process very frustrating. From our survey, we found that our survey respondents started captioning their videos after uploading 30 - 50 videos. The content creator explained her situation about back-captioning problems because she started knowing the captioning process when she already uploaded over 100 videos:

\begin{quote}
    ``...Personally, I did not know I had to caption my videos initially when I became a YouTuber. After uploading over 100 videos, I accidentally watched Rikki's (a deaf YouTuber) video one day, and I realized the importance of having my videos captioned. I would like to claim my channel as a fully inclusive channel. However, if I only start to caption the videos from now on, some people may say something bad in my channel, such as 'your previous videos were not properly captioned.' It literally took me a whole month to caption all past videos without uploading any new videos...''
\end{quote}

In this section, we uncovered 1) caption practices and processes by individual content creators (e.g., the effort of reproducing video transcript for self-captioning), 2) individual content creators' feedback on different caption quality and associated consequences (e.g., caption errors could potentially demonetize the video), and 3) individual content creators' perceptions, general challenges and their awareness on captioning (e.g., back-captioning problems). We further discuss our results and potential design implications for video-sharing platform designers.

\begin{table}[h]
\small
\caption{\red{Summary of Discussion Points}} 
\centering 

\begin{tabular}{|m{3cm}|m{10cm}|} 

\hline 
\textbf{\red{Discussion Themes}} & \textbf{\red{Detail}}\\ 
\hline 
\red{Designing a DHH-friendly Video Recommendation System} & 
\begin{itemize}[left=0cm,topsep=0.2cm]
  \item \red{Create systems with a different set of criteria (e.g., caption availability) to filter out videos.}
  \item \red{Video recommendations should be designed based on video types (e.g., video without any sound, video podcast) to determine whether recommend it if it does not have proper captions.}
  \item \red{Video recommendation systems should allow the audiences to control what to recommend based on accessibility needs.}
\end{itemize}\\
\hline
\red{Minimizing the Effort on Previewing Caption Quality} & \begin{itemize}[left=0cm,topsep=0.2cm]
  \item \red{Leverage audiences' interaction behavior to make caption quality predictions.}
  \item \red{Leverage users' ratings to evaluate caption quality.}
\end{itemize}\\
\hline
\red{Improving and motivating High-quality Community Captions} & 
\begin{itemize}[left=0cm,topsep=0.2cm]
  \item \red{Use machine learning algorithms for specific captioning problems (e.g., irrelevant commentary)}
  \item \red{Monetize community captioning and allocate captioning tasks based on captionists' individual backgrounds.}
\end{itemize}\\
\hline
\red{Presenting Different Levels of Caption Details} & 
\begin{itemize}[left=0cm,topsep=0.2cm]
  \item \red{Detect and extract the abbreviated caption content.}
  \item \red{Display non-speech-based caption content and have different levels of caption density.}
\end{itemize}\\
\hline
\red{Opportunities with Lip-reading} & 
\begin{itemize}[left=0cm,topsep=0.2cm]
  \item \red{Explore lip-reading in the presence of video captions.}
  \item \red{Technologies to track lip movements.}
\end{itemize}\\
\hline 
\end{tabular}
\label{table:DiscussionTable} 
\end{table}

\section{Discussion}
In our results, we answered our research questions by presenting current captioning methods and processes, caption quality and audiences' reactions, caption presentation and improvements (e.g., lip-reading), and challenges of captioning and content creators' perceptions (e.g., back-captioning problems).
In this section, we focus our discussion on 1) designing a DHH-friendly video recommendation system, 2) minimizing the effort on previewing caption quality, 3) improving and motivating high-quality community captions, 4) presenting different levels of caption details, and 5) opportunities with lip-reading \red{(Table \ref{table:DiscussionTable}).}

\subsection{\black{Designing a DHH-friendly Video Recommendation System}}
\red{Video recommendations on YouTube are mainly located in two primary locations---the YouTube home page and the browse page \cite{davidson2010youtube}. Currently, the platform leverages the video metadata (e.g., content creator, publish date) and recommends a video primarily based on the audiences' preference and browsing activities (e.g., watch history and search results) \cite{davidson2010youtube}, but such a video might not be captioned (Section \ref{DHH practice}).} As a result, DHH audiences had to save a captionless video in a playlist and later check for captions (Section \ref{DHH practice}). The delay in captions and the effort to repeatedly check for captions greatly compromised their user experience. To mitigate this problem, we see an opportunity in designing a DHH-friendly recommendation system that operates on a different set of criteria and includes accessibility-related metadata into consideration when developing the recommendation algorithms. For example, the platform could use the availability of captions as a criterion or leverage existing comments and information (e.g., \cite{liu2021reuse,liu2018supporting}) to decide whether or not to recommend a video to a DHH viewer. Furthermore, the time delay in the recommendation should be a function of the characteristics of the target video. For example, a video without any sound should be recommended to the DHH audience right away, while a video podcast should wait until it has been fully captioned. Additionally, we suggest that the DHH audience should have the option to toggle on/off the recommendation delay to fit more customized needs.

\subsection{Minimizing the Effort on Previewing Caption Quality}
In our results, we pointed out the existing concerns from DHH audiences who might click-open a video, only to be disappointed by the poor caption quality after a few minutes through \black{(Section \ref{DHH practice})}. This leads to a further question: how to better predict and present the quality of the captions prior to audiences viewing the video. 

\subsubsection{Leverage Interaction Data to Rate Caption Quality}
From our video analysis, we also learned that not just DHH individuals might need captions, people with auditory processing disorders (V65), non-native speakers (V30), people who are in a situation where having sound is not appropriate (e.g., library) (V64), and young kids who are trying to learn new languages (V148) may also need captions to understand video contents. Therefore, the large number of audiences may provide rich data of their interactions with the video interfaces or reactions towards captions that we can leverage to understand the caption quality. For example, once the system detects the captions have been turned on, it could track where the audience pauses/replays the video or quits to predict the caption at that timestamp may have relatively bad quality. \red{Future systems should further combine existing caption quality prediction metrics (e.g., WER \cite{woodard1982information}, ACE \cite{kafle2017evaluating}) with user's interaction data for better caption quality predictions for different types of videos.}

\subsubsection{Leverage User Score to Rate Caption Quality}
\black{We are aware that many video-sharing platforms, such as YouTube, already form the behavior of using thumbs up/down interfaces to explore users' ratings on videos \cite{siersdorfer2010useful}}. In future work, a video-sharing platform might be able to receive general feedback on the caption quality by prompting the audiences who turned on captions in the video to provide a rating. Therefore, future audiences who leverage captions to understand video content could learn the overall caption ratings from the previous audiences prior to playing the video. \red{Specifically, future work could also explore how the interactive system and user interfaces on social media platforms should be designed to allow audiences to mark the poor caption segment and receive quick clarifications and fixes.}

\subsection{\black{High-quality Community Captions}}
\black{In our results, we mentioned that content creators complained about the caption quality and time delay of community captions (Section \ref{caption practice and processes}). In this section, we further discuss the opportunities of improving and motivating high-quality community captions through machine learning algorithms for captioning problems and monetized community captions.}

\begin{figure}
    \centering
    \includegraphics[width=0.8\columnwidth]{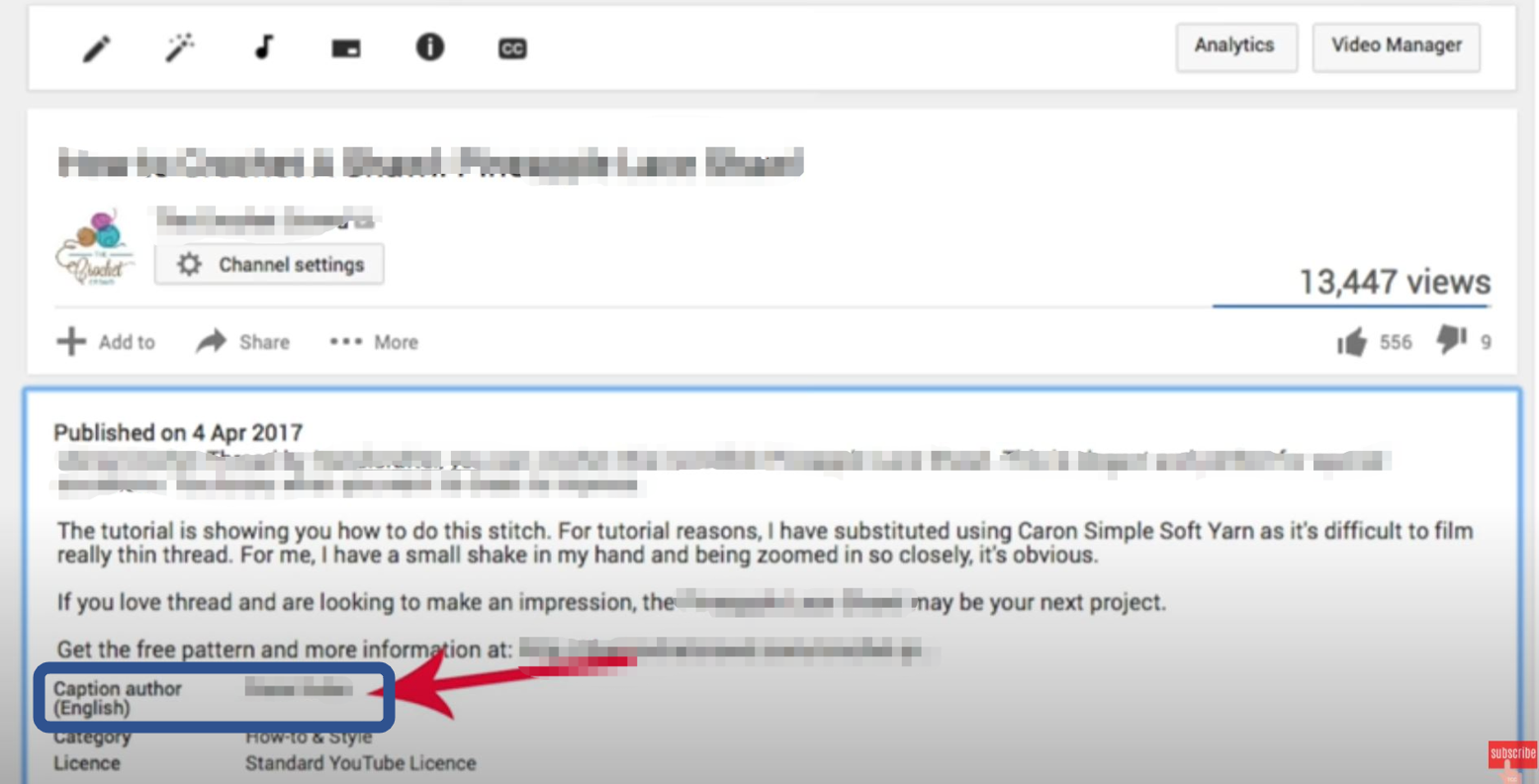}
    \caption{YouTube community-contributed caption credit list in the description of the video.}
    \Description{This figure shows the description of a video. It highlighted Caption author (English) at the bottom left of the figure.}
    \label{fig:communitycredit}
\end{figure}

\subsubsection{\black{Machine Learning Algorithms for Captioning Problems}}
Our findings revealed that many of the individual content creators (46.8\%) leveraged community captions to make their video content accessible to a wider audience, but the community captions are not without problems (e.g., delay, irrelevant commentary) (Section \ref{caption practice and processes}). To improve the crowd caption accuracy, prior research explored machine learning algorithms to combine crowd captions from non-experts and generate more accurate captions \cite{lasecki2017scribe,lasecki2012real}. However, existing research primarily focused on evaluating their algorithms on accuracy and with paid crowd workers. As a potential direction, specialized machine learning algorithms are potentially valuable to specific captioning problems, such as reducing the irrelevant commentary (Section \ref{caption practice and processes}), by combining the captions from different non-expert captionists. \black{Furthermore, future research should also explore the benefits and challenges of certain algorithms in community captioning platforms where most of the non-expert captionists are non-paid volunteers with different background.}

\subsubsection{\black{Monetized Community Captions}}
\black{Currently, the only way of motivating community captions is by listing the contributors' IDs in the video descriptions
(Fig. \ref{fig:communitycredit}), but it does not help generate high-quality community captions.
Some of the content creators suggested that they wish the platform could allow them to use their advertisements revenue income to subsidize the caption contributors.} In V52, the content creator mentioned that he prefers only having one paid community captionist to take the responsibility of captioning each video. \black{To evaluate the caption quality from different community captioning methods, future research could leverage different caption evaluation metrics from existing research \cite{kafle2017evaluating,kafle2019predicting}}. Comparing with third-party captioning companies, having a community captionist who is personally a viewer of a specific content creator's channel or related channels could better caption the related content with their understandings of the channel's domain knowledge (e.g., terminologies in technological products). \black{Therefore, this brings future opportunities to explore the benefit of allocating captioning tasks to community captionists based on community captionists' individual preferences and knowledge}. 



Employing a single paid community captionist may potentially mitigate the delay problem of community captions because the content creator may enable the single captionist to start captioning her video prior to publishing the video. Furthermore, forming a contractual relationship with the community captionist could potentially reduce occurrences of jokes, commentary information, or promotion ads in the captions. However, there might be different practical challenges from the platform level to maintain the connections between community captionists and content creators, which also brings opportunities for developers to design intuitive tools for both groups \cite{liu2019unakite,hsieh2018exploratory}. For example, \black{future work could explore what level of control should the YouTuber grant to the community captionist and how should the platform manage the community captionist and distribute their work.} \red{Additionally, it might be worth exploring ways of motivating high-quality community captions besides monetary rewards. For instance, leveraging a leaderboard mechanism \cite{butler2013effect,sun2015leaderboard} that encourages caption contributors to compete for high user scores.}

\subsection{\red{Levels of Caption Details}}
We touched upon how abbreviations and inadequate presentation of non-speech-based content robbed DHH audiences of an inclusive video-viewing experience (\black{Section \ref{feedback caption quality}}). In this section, we once again bring to readers' attention that DHH audiences desire more details in the caption, but how to address this need remains a future research question.
\subsubsection{\red{Abbreviated Caption Content}}
In our results, we included examples of where captions used [Joke] or [Music] to represent the actual joke or the music. We also concluded that DHH audiences have a negative feeling towards just inserting [Joke] in the captions to represent the actual content. We found the key reason behind this problem is that DHH audiences feel they are not treated equally on video content (P13). 
We also learned that some content creators chose to use [Swear Word] instead of actually displaying the swear word in the captions, even if they did say that word in the video, to prevent getting demonetized or yellow flagged. \textit{``I do not want other people to decide what I should know or not from the video, whatever the person said in the speech, just put whatever in the caption,''} said P5. This situation also applies to racist and adult-only content. P9 continued: \textit{``I am an adult, and I do not mind having adult-only content in the caption, whatever is said in the video, just show them in the caption. If it is adult-only, you should add a notification once audiences opened your video to prevent kids from watching it, not modifying the caption.''} \black{Therefore, future research should take accessibility needs into considerations when designing filters of different contents to make sure the equity and inclusion \cite{chadwick2016digital} of the video contents to DHH audiences.}

\subsubsection{Non-speech-based Caption Content}
For audio-speech information, captionists could make sure the DHH audiences get the same level of content by transcribing the speech word-by-word in the caption. However, some auditory contents are hard to transcribe if they are not speech-based. For instance, describing music and background sounds with appropriate levels of detail would be a challenge. Too few details might throw the audience off, but too many details would take many lines of text. In fact, from our video analysis and interviews, we found most of our DHH audiences prefer having closed captions that do not have over two lines (P1 - P13, V3), and some complained that too much content might stack up on top of each other in captions (V93). \red{Future research could explore different algorithms to identify features of the non-speech sound (e.g., \cite{sporka2006non,cowling2002recognition})} that are most descriptive to DHH audiences and present them within a proper length limit. Additionally, DHH audiences' demand for levels of caption details might vary. One should consider giving DHH audiences an option to switch between a `verbose' mode and a `concise' mode.


\subsection{\black{Opportunities with Lip-reading}}
DHH interviewees have stressed the importance of leveraging lip-reading to supplement their understanding of video content in addition to reading captions (Section \ref{feedback caption quality}). \black{In this section, we outlined two research directions that we can further pursue with lip-reading.} 
\subsubsection{\black{The Role of Lip-reading for Online Videos}}
\black{According to Richard Mayer's theory of dual encoding, multiple sources of visual information could compete for attention in the visual channel, which could potentially result in a distraction \cite{mayer2005cognitive}. Therefore, the effect of lip-reading in the presence of video captions remains a research question for further exploration. More specifically, future research could investigate whether lip-reading serve as a complement for captions or a replacement for captions and DHH audience's behaviors of allocating their attention between reading captions and reading lips.}

\subsubsection{\black{Make Lips Visible}} 
\black{For DHH audiences who have grown accustomed to reading lips and captions at the same time, they might find the viewing experience incomplete if the speaker in the video unintentionally occludes his or her mouth (Section \ref{feedback caption quality}).} For example, many content creators show their face at an angle that is difficult for DHH viewers to see the mouth (Fig. \ref{fig:lipreading}). One direct solution to make lips more visible to the DHH viewers is to have the speaker equip a wearable camera \cite{li2019fmt,kianpisheh2019face} or explore earable sensing systems \cite{sun2021teethtap} that track the lip movement. However, such an approach would be very cumbersome and costly to individual content creators. Another possible solution is to leverage the latest computer vision techniques to reconstruct the speaker's entire face from limited inputs. For example, Elgharib et al. \cite{elgharib2020egocentric} proposed a method that generated the speaker's front face view from only the side face view, which was provided by a low-cost wearable egocentric camera. \black{Future research should explore how DHH audiences would react to the adoption of such accessibility technologies by content creators.}

\section{Limitation}
We acknowledge several limitations in our study that may have implications on how to interpret our results. We recruited both interviewees and survey respondents through social platforms, such as Facebook Group, Reddit, and Twitter. However, we may not cover the sample who do not use these social platforms. Furthermore, our survey may only attract those content creators who are willing to take surveys. In our survey study, most respondents are content creators who have less than 10k subscribers, and we did not have survey respondents who have over 1 million subscribers. This may lead to different practices and understandings of captioning approaches. 
In our video analysis and survey, we only covered videos and content creators who speak English or have English captions. However, captioning practices may vary if the content creators speak a language and caption consumption practices might be different across different cultures \cite{li2021choose}. \black{In terms of using YouTube videos as a data source, beyond its benefits that we described in the methodology section (e.g., \cite{anthony2013analyzing}), there may also be tradeoffs and risks (e.g., privacy) on using YouTube videos \cite{klobas2018compulsive,myrick2015emotion,spinelli2020youtube}. We look forward to exploring more about how the risks and tradeoffs of using YouTube videos affect the DHH population and individual content creators' experiences with captions.}


\section{Conclusion}
In our work, we conducted an interview study with 13 DHH YouTube viewers, a video analysis with 189 videos on Youtube, and a survey study with 62 content creators, to explore the knowledge gap between DHH audiences and individual content creators on captioning. We introduced DHH audiences' practices, perceptions, and challenges on watching captioned videos by individual content creators, such as enhancing lip-reading qualities and avoiding using abbreviated terms in ``[]''. We also revealed the problems of different captioning methods by individual content creators (e.g., community captionists are unwilling to help captions for videos that they are not familiar with) and showed their perceptions of captioning and associated challenges (e.g., back-captioning problems). \black{We extended our findings through the discussion of developing a DHH-friendly video recommendation system, minimizing the effort on previewing caption quality, improving and motivating high-quality community captions, presenting different levels of caption details, and opportunities with lip-reading}. Overall, our work provides an understanding of the existing gaps between individual content creators and the DHH audiences on video captions.
We believe our design suggestions could shed light on the creation, interpretation, and presentation of captions generated by individual video content creators and provide design implications to video-sharing platform designers.

\begin{acks}
This work was supported in part by the Natural Sciences and Engineering Research Council of Canada (RGPIN-2016-06326). 
\end{acks}

\bibliographystyle{ACM-Reference-Format}
\bibliography{main}


\end{document}